\documentclass[%
 reprint,
 amsmath,amssymb,
 aps,
]{revtex4-2}

\usepackage{graphicx}
\usepackage{xcolor}
\usepackage{dcolumn}
\usepackage{bm}

\usepackage{hyperref}

\begin{document}

\preprint{APS/123-QED}

\title{Self-balancing luminescence kinetics of charged excitons in WS$_2$ monolayer}


\author{A.I. Veretennikov}%
\email{veretennikov.a@mail.ioffe.ru}
\author{A.I. Galimov}%
\author{M.M. Glazov}%
\author{T.V. Shubina}%
\affiliation{Ioffe Institute, St. Petersburg, 194021, Russia}%

\begin{abstract}
Two-dimensional tungsten-based transition metal dichalcogenides exhibit a rich emission spectrum with distinct lines of  exciton and biexciton states, neutral and charged, which are typically considered as independent entities. Here, we present a study of the emission kinetics of a WS$_2$ monolayer, demonstrating that these states together form a single interacting ensemble, with their population and depletion mutually coordinated with temperature. Dark trions make a long-lived contribution to the bright trion on a time scale of 200-300 picoseconds, and replenishment from dark states promotes intense luminescence of charged biexcitons. The developed self-consistent rate equation model reproduces the temperature dependence of the kinetics well, opening the way to controlling exciton states in such materials.
\end{abstract}

\maketitle


\section{\label{sec:level1}Introduction}

\begin{figure}[h]
	\centering
	\includegraphics[width=0.9\linewidth]{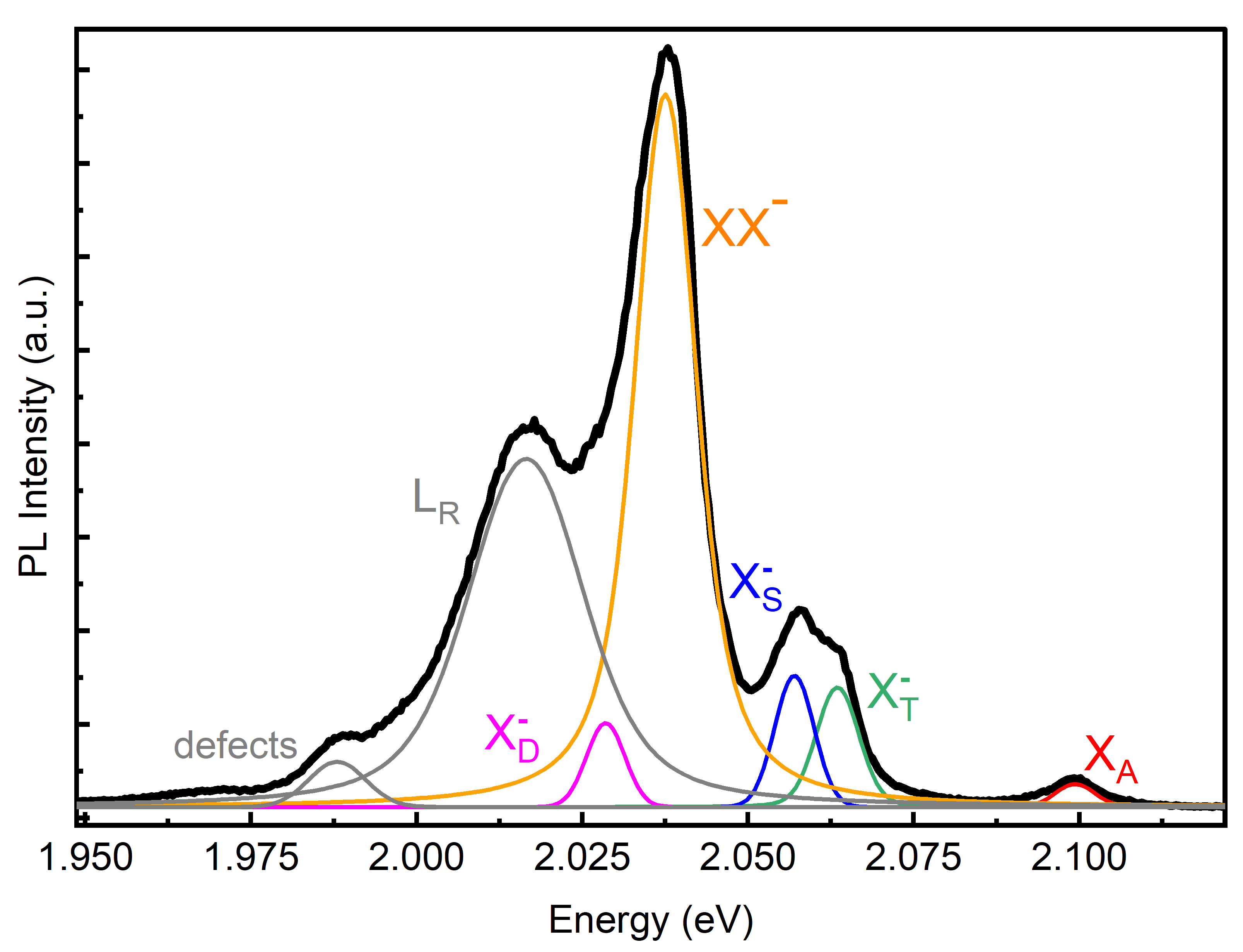}
	\caption{$\mu$-PL spectrum of a monolayer WS$_2$. The black curve shows experimental data; the color peaks represent decomposition of the overall spectra into individual excitonic species contributions with the following notations: $X_A$ - bright exciton, $X_{S/T}$ - intra- and intervalley bright trion, $XX^-$ - charged biexciton, $X_D^-$ - dark trion, $L_R$ - phonon replicas.}
	\label{fig1}
\end{figure}

Transition metal dichalcogenide (TMD) monolayers have become one of the most extensively studied systems in the family of two-dimensional (2D) semiconductor materials due to their unique physical properties, which hold promise for many applications \cite{Novoselov2016,Pu2018,Wang2018, Yu2016,Liu2019QIS}. In tungsten-based monolayers, called ''darkish'', the inversion of the conduction subbands makes the exciton ground state optically inactive \cite{Zhang2015,Echeverry2016,Malic2018,Wang2017}. This electronic structure allows observing a rich photoluminescence (PL) spectrum which demonstrates, along with the bright excitons, a number of pronounced lines related to the dark exciton states and charged complexes, a typical example of WS$_2$ monolayer PL is shown in Figure \ref{fig1}. 

The distinct  interpretation of these lines is of great importance, for example, for opto-valleytronic \cite{Xu2014}, transport~\cite{chernikovChapterThreeExciton2023} and quantum photonics \cite{Parto2021} applications. The nature of the most pronounced peaks in the PL spectra has been discussed in a number of papers \cite{Plechinger2016,Bao2020,Pei2023}.  Experimentally, the origin of each emission line is usually determined from its spectral position, kinetics, evolution in external magnetic or electric fields, and optical orientation experiments. Interestingly, already at moderate excitation powers, the spectra are dominated by multiparticle and dark states rather than by bright excitons. The origin of these lines -- whether they are related to the charge-neutral or charged entities\cite{Nagler2018,Koperski2019} or to localized states\cite{Jadczak2017,Vaclavkova2018} -- has been debated in the literature with the consensus regarding the main emission peaks reached only recently\cite{Chen2018,Yang2022,Chatterjee2022,Ren2023,Zhu2025}.
 
Furthermore, despite the fact that the influence of dark states on the luminescence intensity and kinetics has been investigated in many works \cite{Zhang2015,seligDarkBrightExciton2018,seligUltrafastDynamicsMonolayer2019,Li2019,Liu2019,Brem2020,Mondal2022}, the origin of observed lifetimes in the range of hundreds of picoseconds for charged exciton states \cite{Robert2016,Liu2019,Li2019,Zipfel2020,Pei2023} remains incompletely explained, although it may indicate replenishment from dark states \cite{Arora2020,fangControlExcitonRadiative2019,RosatiSidebands2020}. Clearly, the intertwined population dynamics of different exciton states in W-based TMD monolayers requires a more systematic consideration, as this is a key issue for transport applications involving long-lived excitons~\cite{Mueller2018,wagnerNonclassicalExcitonDiffusion2021,rosatiDarkExcitonAntifunneling2021,chernikovChapterThreeExciton2023,malicExcitonTransportAtomically2023}.

\section*{Experimental results}

\begin{figure*}[t]
	\centering
	\includegraphics[width=\linewidth]{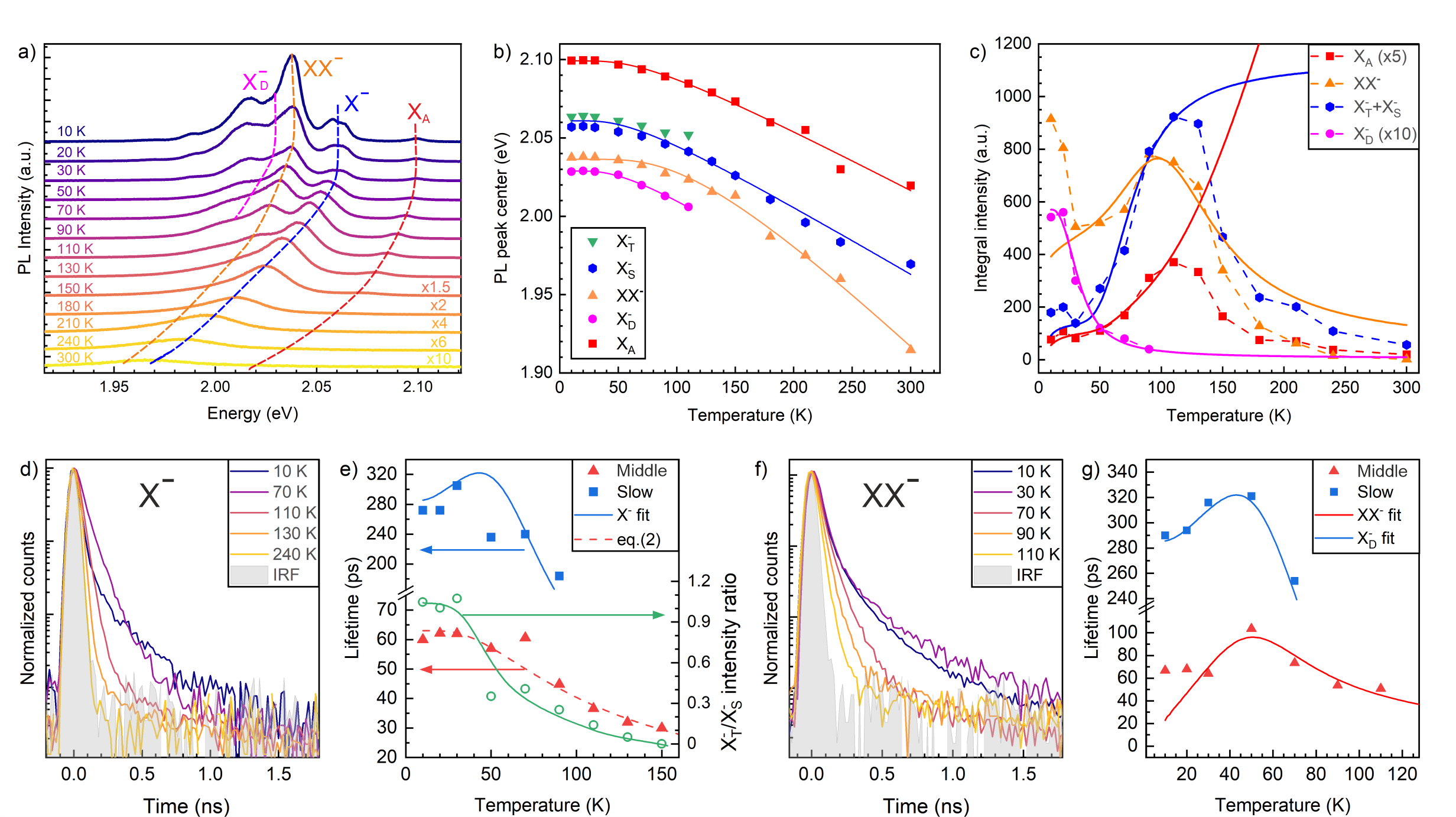}
	\caption{ a) $\mu$-PL spectra of the WS$_2$ monolayer measured at different temperatures. b) Temperature dependencies of peak energies and c) integrated PL intensities (color symbols) extracted from the spectra in panel (a). Solid lines in (b) show the fit by the model of Ref.~\cite{PhononCotangent}, while and solid lines in panel c) show the fit by the proposed model.  Fitting parameters are given in Tables~1 and 2 in SM. d) Measured PL decay curves for the trion line at different temperatures. e) Characteristic times extracted from the decay curves of the $X^-$ line: left axis — middle times (red triangles) and slow lifetimes (blue squares — experiment, blue solid line — model fit); right axis — the integrated intensity ratio of the $X^{-}_S$ and $X^-_T$ lines in the trion doublet at different temperatures. f) Measured decay curves for the $XX^-$ line at different temperatures. g) Middle (red triangles) and slow (blue squares) characteristic lifetimes obtained from the decay curves of the $XX^-$ line. Solid lines show the fit using the proposed model. The shaded area in d) and f)  is the IRF obtained by detecting the backscattered laser pulse.}
	\label{fig2}
\end{figure*}

 The rich photoluminescence (PL) spectrum of non-encapsulated WS$_2$ monolayer at $T=10$~K shown in Fig.~\ref{fig1} consists of several distinct peaks as demonstrated by its decomposition into several Voigt functions. In accordance with previous studies \cite{Vaclavkova2018,Zinkiewicz2020,Robert2021,Yang2022,Ren2023},  we can identify a relatively weak peak of the bright exciton $X_A$ (red peak at $2.1$~eV), a doublet of the negatively charged exciton ($X^-$-trion)  (blue and green peaks corresponding to the split singlet $X_S^-$ and triplet $X_T^-$ trion states\cite{Courtade2017,Zipfel2020,durnevExcitonsTrionsTwodimensional2018}), and a charged biexciton $XX^-$ (orange peak) dominating the PL spectrum at low temperature. The biexcitonic nature of the $XX^-$ complex was confirmed through power-dependent PL measurements demonstrating its superlinear behavior, see Supplementary Materials (SM) for details.  For a reliable approximation of spectra and kinetics at temperatures below 90 K, we found it necessary to include a weak feature near the  $XX^-$ line that, in agreement with Refs.\cite{Zinkiewicz2020,Chen2018,He2020}, corresponds to the dark negative trion $X_D^-$  with noticeable PL intensity. A broad feature $L_R$ consisting of several interlapping phonon replicas and the defect luminescence is observed in the energy range of $1.95-2$~eV.

 Remarkably, despite the characteristic emission from the dark trion state, the PL signal from the dark neutral exciton $X_D$, expected at 2.045–2.055 eV, was hardly detectable in our experiments. This is consistent with previous reports showing that $X_D$ is observed only near charge neutrality\cite{Chen2018,Barbone2018,Yang2022,Ren2023,Lagarde2024}. In our case, when the monolayer was obtained from a negatively-doped bulk crystal, the trion PL is significantly stronger than the neutral exciton PL. Therefore, we expect the $X_D$ emission to be correspondingly weaker than $X^-_D$. A similar effect was observed in previous studies with TMD monolayers \cite{Chen2018, Paur2019, Zinkiewicz2020}. 

\begin{figure}[t]
	\centering
	\includegraphics[width=\linewidth]{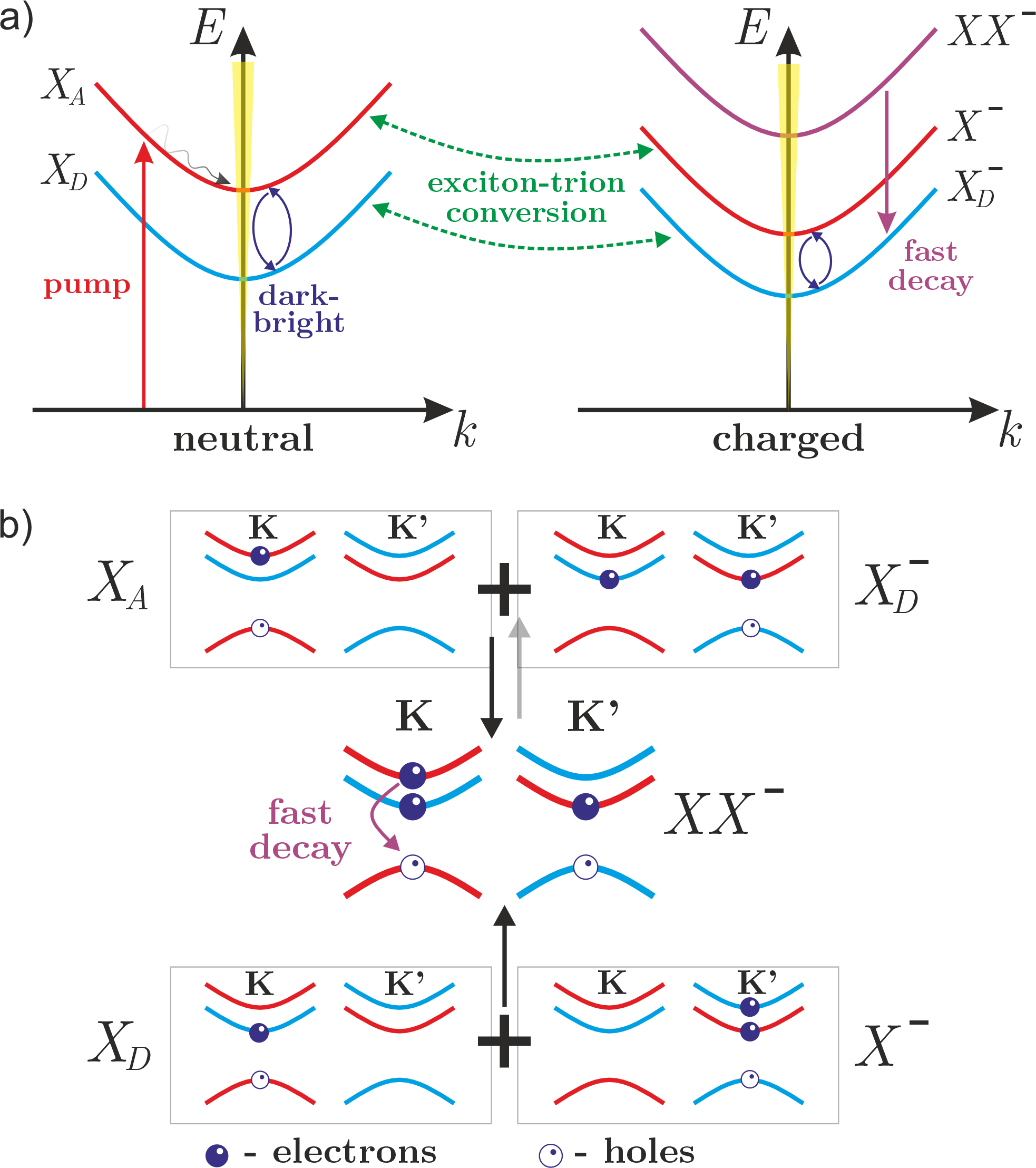}
	\caption{ 
    a) Scheme of excitonic states taken into account in our model (bottom of all band centered in $K-K'$ valley). Respective excitation and re-population processes are shown by arrows. b) Considered mechanisms of $XX^-$ formation (black arrows) and dissociation (gray transparent arrow), purple arrow depicts fast radiative decay of the $XX^-$. Red (blue) parabolas represent spin-up (-down) sub-bands. }
	\label{fig5}
\end{figure}

Figures \ref{fig2}a-c show the temperature variation of the $\mu$-PL data. In addition to the overall  redshift of the spectrum related to the temperature-induced band-gap reduction\cite{PhononCotangent}, see Fig.~\ref{fig2}a and b, a significant redistribution of the relative peak intensities is observed. At elevated temperatures, the $XX^-$ peak loses its dominant role and practically disappears above 110 K. Instead, the PL signal becomes dominated by trion emission up to room temperature. 

Notably, the intensities of all peaks (except $X_D^-$) increase almost simultaneously in the temperature range of 50–110 K as illustrated in Fig.~\ref{fig2}c. The subsequent decrease in emission intensity at temperatures above 150 K is likely due to the acceleration of nonradiative recombination. Therefore, further discussion will focus on temperatures in the range of 10–130 K.

The closeness of the temperatures at which the intensities of individual peaks increase  indicates the interconnected dynamics of the complexes, which requires detailed analysis. To determine the role of excitation transfer between different exciton states in the observed PL enhancement, we used temperature-dependent  spectroscopy of time-resolved photoluminescence (TRPL), measuring the temporal dynamics of each line observed in the spectra at different temperatures (experimental details are provided in SM).

For reasonably accurate TRPL data approximation, we used up to three exponential functions to fit the decay curves and extract characteristic decay times corresponding to fast, middle, and slow decay processes (the fitting procedure is described in SM). For the bright exciton transition, the PL decay curves are largely monoexponential with a short radiative lifetime; long-lived components were not detected over a dynamic range of almost three orders of magnitude (for the $X_A$ decay data see SM). This behavior is consistent with previous studies and can be explained by the large energy separation between the bright exciton and the dark exciton amounting to 45-55 meV in this material system\cite{Wang2017, Li2019, Zinkiewicz2020}, and, accordingly, to the negligible probability of their repopulation as a result of the dark state activation.

In contrast to the bright exciton, the bright trion $X^-$ exhibits  long-lived components  (Fig.~\ref{fig2}e), consistent with activation from dark states. At temperatures below 150 K, the decay curves can be satisfactory fitted only using three exponential components: fast ($<15$~ps), middle (30-60 ps) and slow (180-320 ps). The colored symbols in Fig. \ref{fig2}e represent the evolution of the middle and slow radiative lifetimes as a function of temperature, obtained from decay curves measured at the center of the trion $X^-$ doublet. 

The PL kinetics for the $XX^{-}$ line is shown in Fig.~\ref{fig2}f.  Here we also observe a slow component on time scales of 250–300 ps (blue squares in Fig.~\ref{fig2}g). This can be explained by the admixture of the dark trion $X_D^-$ emission due to the strong overlap of this line with $XX^{-}$ at temperatures below 70 K. Indeed, these slow lifetimes almost coincide with the slow times in the bright trion emission, caused by the interplay with the dark trion state. In the $XX^{-}$ PL, along with components with fast and slow characteristic decay times, we observe components with middle times of 60–100 ps. This intermediate-lived component contributes 40\% to the total emission at 10 K. The simultaneous presence of fast and middle components in the TRPL decay was not observed in previous works on biexciton recombination in WS$_2$ \cite{Nagler2018,Chatterjee2022}. Given the many-body nature of the $XX^{-}$ complex, this middle component could be inherited from trion states involved in charged biexciton formation \cite{Chen2018,Barbone2018}.

It should be noted that, typically, activation from underlying dark states results in long-lived components in the luminescence of bright states which have naturally shorter lifetimes \cite{Labeau2003, Zhang2015, Zinkiewicz2020, Eliseyev2021}. A similar effect leading to the appearance of a slow component has also been observed in ''darkish'' 2D materials \cite{Zhang2015, Robert2016, Liu2019, Li2019, Brem2020, Zipfel2020, Pei2023}.

The observation of slow and middle components in the PL decay curves suggests the presence of both (i)~efficient activation processes that can occur from a dark to a bright state or from a neutral to a charged state, and (ii)~excitation transfer channels via bimolecular formation of biexcitons followed by their thermal dissociation. The high intensity of the $XX^-$ with complex temperature-dependent kinetics indicates the efficient transformations between excitons specifically in this way.  In this framework, for analysis of the experimental PL kinetics bimolecular formation processes are no less important than direct conversion processes between separate complexes.

\section*{Model}

To obtain in-depth insight into the processes underlying the PL, we develop a kinetic model describing population transfer between relevant species in the system.

A schematic representation of the exciton complexes with the corresponding optical transitions is shown in Fig.~\ref{fig5}a. Here, we consider only the following states: the bright exciton $X_A$ and its dark state $X_D$, the bright negative trion state $X^-$ and its dark state $X_D^-$, and the charged biexciton state $XX^-$. In our model $X_D$ state effectively accounts for both intravalley (spin-forbidden and momentum allowed) and intervalley (spin-allowed momentum-forbidden) neutral dark excitons. The formation scheme of $XX^-$ is shown in Figure \ref{fig5}b. It emphasizes the importance of the dark trion as a reservoir for $XX^-$ and as the final product of its dissociation. In this model, we neglect the fine structure of the bright trion, since its splitting is significantly smaller than the typical binding energy and the bright-dark splitting for both the exciton and the trion. This does not preclude further consideration of the exchange interaction effect on the details of the observed kinetics.

Numerical calculations for modeling the self-consistent population kinetics of exciton states, taking into account all essential processes, are based on the rate equations in a general form
\begin{equation}\label{eq1}
\frac{d}{dt}N_i = -\sum_j\kappa_{ij}N_j-\sum_{k\not=i}w_{ik}N_iN_k,
\end{equation}
where $N_i$ denotes the population of the $i$-th complex where the subscript $i$ runs, in agreement with aforesaid, through $A$ for $X_A$, $T$ for $X^-$, $D$ for $X_D$, $DT$ for $X_{D}^-$ and $XX$ for $XX^-$ (cf. Fig.~\ref{fig5}).  The first term on the right side of equation (\ref{eq1}) describes the single-particle processes, including (i)~recombination with the rate $\kappa_{ii} = \tau_i^{-1}(T)$, and (ii)~phonon-mediated dark-to-bright\cite{Labeau2003,Arora2020,Chatterjee2022,Mourzidis2025} or neutral exciton-to-trion\cite{Jones2016,Ayari2020} and vice-versa conversion processes, where the effective rates are related by $\kappa_{ij}/\kappa_{ji} = -\exp(\Delta_{ij}/kT)$. Here $\Delta_{ij}$ is the energy separation between the $i$ and $j$ states, $k$ is the Boltzmann constant. We assume that the resident electron density is much larger than the density of photogenerated excitons that allows us to neglect the depletion of resident electrons in the course of trion formation.
These conversion processes are illustrated by colored arrows in Fig. \ref{fig5}a: vertical arrows for recombination, dark cyclic arrows for dark-bright interplay and green arrows for exciton-to-trion conversion.

 The last term in Eq.~(\ref{eq1}) is included to account for the nonlinear processes with the population transfer from one complex to another, namely, the formation of charged biexciton as a result of the exciton and trion fusion. The corresponding schemes are depicted in Figure \ref{fig5}b: we assume the $XX^-$ arises from $X_A+X_D^-$ or $X^-+X_D$ combinations, and the pair of complexes $X_A+X_D^-$ is the most probable outcome of the $XX^-$ dissociation. Hence, the decay of $XX^-$ results  in the refilling of the excitons and trions with the corresponding rates denoted as $\kappa_{i,XX}$ ($i=A,DT$). In the pulsed excitation regime (as in our experiment), the generation rate is introduced through the initial conditions for the bright population $X_A$. The coefficients \(\tilde\gamma,\ w_{ij}\text{ and }\kappa_{ij}\) depend on the thermal energy $kT$.

The complete system of rate equations and the corresponding functional relations are given in SM. With the exception of the dark exciton $X_D$ lifetime, the $X_A$ bright-dark splitting, and bimolecular formation constant for the $XX^-$, microscopic parameters such as the energies and rates of transitions were taken from experiment or previous theoretical works \cite{Robert2016,Palummo2015,Ayari2020,Lagarde2024}. Using this formalism, we  have modeled the decay curves of all exciton complexes under study.

\section*{Analysis and discussion}

The developed model allowed us to model the results of the TRPL measurements. Remarkably, we simulated TRPL decay curves for all complexes simultaneously, using a single set of parameters from present PL measurements and previous works (see Tab.~\ref{tabS5} in SM). Since the decay kinetics of the $X_A$ exciton are extremely fast (both observed and calculated), we limit further discussion to data for the $X^-$ trion and the charged $XX^-$ biexciton.

\subsection*{Trion $X^-$ kinetics}

Figure~\ref{fig2}e shows the decay times of the $X^-$ trion population extracted from the curves in Fig.~\ref{fig2}d at different temperatures as described above (points) and the calculated times found by the analogous fitting of the numerically simulated $N_{T}(t)$ curves.

The simulated decay curves were satisfactorily fitted by two exponential functions with characteristic decay times in the range of 5-10 ps (fast) and 200-320 ps (slow). The simulations of kinetics at different temperatures revealed two main processes governing trion kinetics and population: exciton-to-trion conversion and dark-to-bright trion transition from $X_D^-$. It is reasonable to attribute the slow component to the thermal re-population of the $X^-$ state from the dark state $X^-_D$. At low temperatures ($\sim$10~K), simultaneously with radiative recombination, optically generated excitons rapidly ($\sim$3~ps\cite{Ayari2020}) transform into negative trions, followed by a slower ($\sim$10~ps\cite{Lagarde2024}) relaxation to the dark trion state. This fast cascade results in a relatively low population of bright trions at temperatures below 30 K. At higher temperatures ($\sim$70 K), thermal equilibrium is reached in both processes, but exciton-trion conversion still occurs much faster. An increase of the temperature also stimulates the dark-to-bright transition from dark trion $X_D^-$, leading to emergence of additional slow component in $X^-$ kinetics (blue data in Fig.~\ref{fig2}e) and rise of integrated bright trion PL (blue solid line in Fig.~\ref{fig2}c). It also shortens the effective lifetime of the dark trion $X_D^-$. Thus, the experimental behavior of the temperature-dependent fast and slow components of the bright trion recombination in the WS$_2$ monolayer PL agrees well with the model description.

To clarify the origin and kinetics of the middle component, we approximate the temperature evolution of the recombination time as in a three-level model involving a ground state and two split excited states, originally used to study the splitting of bright and dark excitons in low-dimensional structures \cite{Rodina2016, Golovatenko2022}:
\begin{equation}\label{eq2}
	\tau_m = \gamma_0^{-1}\cdot\tanh\left(\frac{\Delta_{ph}}{2kT}\right).
\end{equation} 
Here, $\gamma_0$ is the rate of excitation relaxation between two split states at $T=0$~K, $\Delta_{ph}$ is the phonon energy required to overcome the gap between them, and $k$ is the Boltzmann constant. From the fitting we obtained \(\Delta_{ph}=13.8\pm0.5\) meV and $(\gamma_0^{-1}=62)$~ps$^{-1}$. One of the possible mechanisms responsible for the middle lifetime may be unaccounted population transfer involving the higher-energy $Q$-valley state. The obtained value of $\Delta_{ph}$ also agrees well with the energy difference between excitons in the $K-K'$ band and the $K-Q$ band \cite{Peng2019}. According to recent theoretical studies \cite{Lo2021}, the transition between these excitons via the phonons should be significantly slowed down compared to the direct intravalley transition in comparison with intravalley $K-K'$ relaxation through the $K_3$ phonon\cite{Ren2023,Lagarde2024}. However, the formation process of the $K-Q$ trion remains unclear, and the influence of this high-energy state on the observed exciton complexes is still debated.

\subsection*{Charged biexciton $XX^-$ kinetics}

\begin{figure}
    \centering
    \includegraphics[width=0.9\linewidth]{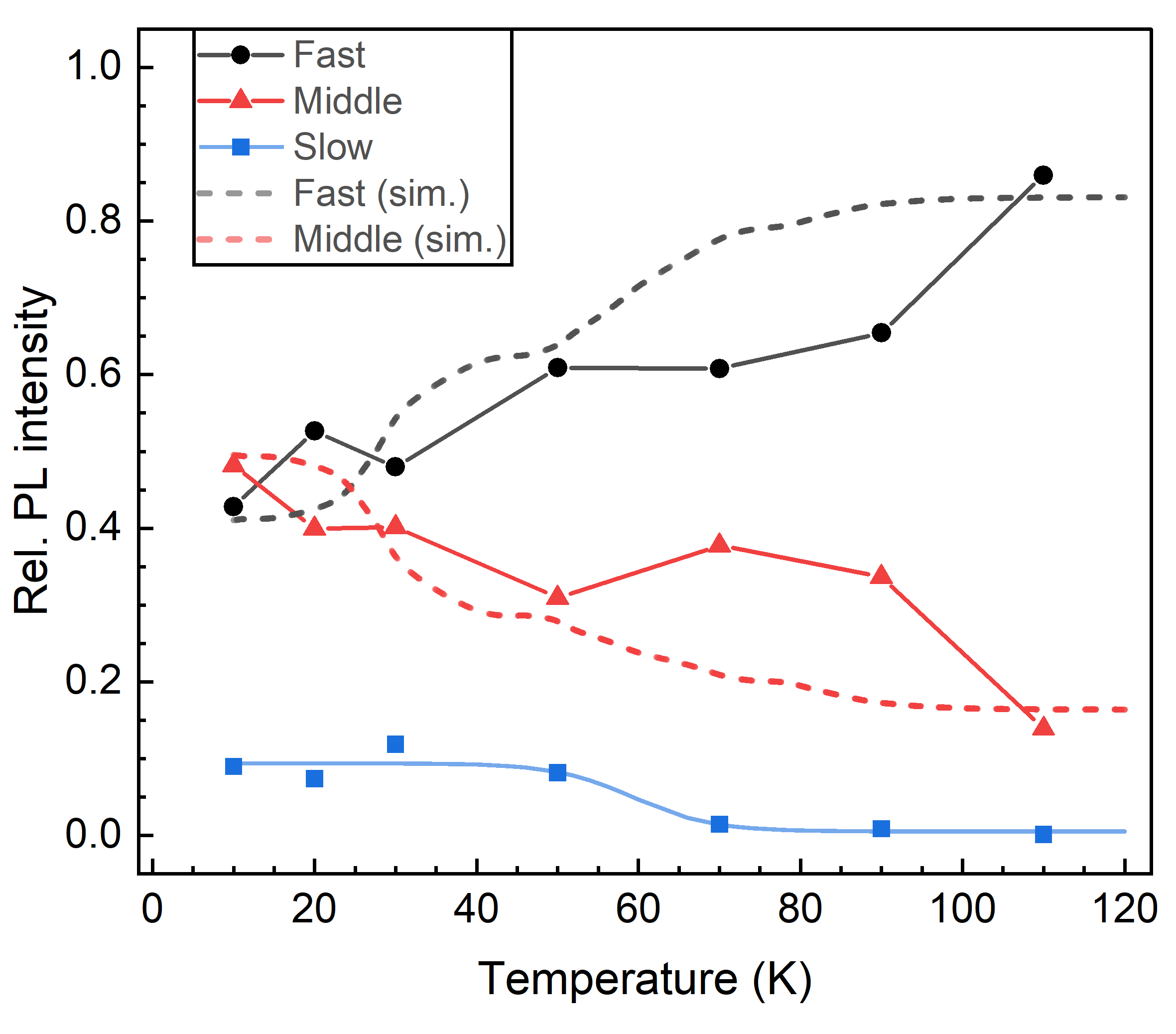}
    \caption{The relative contributions of fast (black), middle (red) and slow (blue) to the total luminescence of $XX^-$ complex. Solid lines with symbols represent the data extracted from TRPL curves fitting, dash curves illustrate impacts obtained from model. Light-blue line is shown to guide the eye.}
    \label{XX_rel}
\end{figure}

The decay times of the charged biexciton dynamics extracted from simulated decay curves are shown in Fig. \ref{fig2}g by solid lines. 
The simulation shows that the fastest lifetimes in the $XX^-$ kinetics (not shown) are close to the lifetimes calculated for the $X_A$ exciton. This result follows from the fact that the most probable $XX^-$ decay channel is the recombination of an electron and a hole with parallel spins in the same valley (purple arrow denoted as ''fast decay'' in Fig. \ref{fig5}b). The contribution of the slow component to the total luminescence intensity of $XX^-$ is about 10\% at 10~K and decreases at higher temperatures, almost vanishing at $T>$70~K (see blue dots in Fig.~\ref{XX_rel}). The temperature dependence of the slow lifetimes follows well that of $X_D^-$ discussed in the previous subsection (cf. blue symbols in Figs. \ref{fig2}g and \ref{fig2}e). This matching challenges the previous assumption that the slow component is due to emission from defect states \cite{Plechinger2016,Jadczak2017,Vaclavkova2018}.\\

For the middle lifetime (red symbols in Fig. \ref{fig2}g), our model accurately reproduces the experimental non-monotonic dependence for temperatures above 30 K with a peak at around 50 K. The simulations show that the biexciton population and kinetics are mainly determined by its bimolecular formation. At the lowest temperatures, the $XX^-$ complex is formed from the exciton $X_A$ and the dark trion  $X_D^-$ as shown by the upper part of Figure \ref{fig5}b, and corresponding kinetics inherits the fast lifetime of $X_A$ and the slow lifetime of $X_D^-$. With increasing temperature, the $XX^-$ decay rate decreases due to the increasing influence of $X_D^-$ and also due to the escape of the exciton states from the light cone \cite{Robert2016,Palummo2015} (see SM). This slowdown persists until the onset of thermal decomposition of $XX^-$ into dark exciton $X_D$ and bright trion $X^-$, accompanied by a simultaneous rapid decrease in the lifetime of the dark states, which is reproduced in our model, see the drop of the blue line in Figure \ref{fig2}g. 

Proposed mechanism is additionally supported by temperature dependence of relative impacts of fast and middle components in total PL intensity, shown in Fig.~\ref{XX_rel}. At 10~K contributions of both fast and middle components are almost equal 45\%. Observed proportion indicates that slowed impact from dark states influences the total PL intensity as much as the fast recombination. With increasing temperature contribution of the middle lifetime drops to 10\% at 110 K due to both enhanced dissociation and shortened lifetimes of dark states. At the same time, the fast component, in agreement with the proposed mechanism, exhibits an increase. Experimental trend is approximately reproduced by simulations that take into account the excitation transfer between different states.

A discrepancy between the experimental and simulated lifetimes of $XX^-$ at temperatures below 30 K can be due to the uncertainty in the middle $X^-$  lifetimes. Since the trions participate in the formation of biexcitons via fusion with the dark exciton $X_D$ (lower part of Figure  \ref{fig5}b), an increase in the $X^-$ population at low temperatures leads to the growth in the $XX^-$ population and the appearance of a decay lifetime governed by the $X_D$ and the $X^-$ kinetics through the last term in Eq.~(\ref{eq1}) (see explicit form in SM). This is consistent with the approximation of the integrated PL intensity in Figure \ref{fig2}c: at temperatures below 30 K, the intensities of $XX^-$ and $X^-$ differ from the simulated values by approximately a factor of 2, while the intensities of the $X_A$ and $X_D^-$  are in good agreement with experiment. Thus, an increase in the population of $X_D^-$ at temperatures below 30 K leads to an additional increase in $XX^-$ PL unaccounted for in the model.

We have performed additional consistency tests for the model. It is important to note that the temperature dependencies for the $XX^-$ decay cannot be reproduced when assuming that the $XX^-$ line is that of a neutral biexciton instead of a charged one: in this case, the biexciton binding energy would exceed 60 meV while experiments and calculations report much lower values $\sim 25$~meV~\cite{conwayDirectMeasurementBiexcitons2022,mostaaniDiffusionQuantumMonte2017}. Simulations without taking into account the influence of dark trions will yield a monotonic increase in the lifetime in the considered temperature range otherwise, because of the high binding energy of the dark neutral exciton $X_D$ the peak in the temperature dependence of the lifetimes would shift towards higher temperatures. Moreover, the formation of the $XX$ complex from neutral excitons will lead to a significant decrease in the emission intensity \cite{Chen2018,Paur2019}. In contrast, when considering the $XX^-$ complex together with negatively charged trions, we obtain a practically ideal fit of the experimental data. Remarkably, this is done with the assumption of the charged exciton binding energy (with respect to the $X^-$) of 22.6$\pm$0.3 meV, which is in good agreement not only with the energy separation we measured, but also with the data of previous studies \cite{kylanpaa2015,
 zhangExcitedBiexcitonsTransition2015, kiddBindingEnergiesStructures2016, Szyniszewski2017, mostaaniDiffusionQuantumMonte2017, Chen2018, Barbone2018}.

\section*{Conclusions}
 In this work, we examined the temporal dynamics of the complicated emission  spectrum of a negatively doped WS$_2$ monolayer, which includes distinct lines of charged excitons (trions, both bright and dark) and a charged biexciton.  Using temperature-dependent time-resolved PL spectroscopy, we clearly demonstrate interrelated  and self-balanced population dynamics of different exciton states. Three scales of decay times, fast, middle, and slow, were identified. In particular, the long-lived component ($\sim 300\ \text{ps}$) observed in the decay of bright trions arises from their thermally activated formation from dark trions $X_{D}^{-}$. This represents a mechanism for the previously unexplained slow $X^-$ dynamics.
 
Furthermore, we have established that the strongest PL line at low temperature is the emission of  the charged biexciton $XX^{-}$, a complex that can form from two constituents: a neutral exciton and a charged trion. The state $XX^{-}$ exhibits a complicated decay, in which, at low temperature, a well-pronounced component with the middle lifetime also reproduces  the emission features of  the dark trion $X_D^{-}$. Overall, the temperature-dependent changes in the $X^{-}$ and $XX^{-}$ lifetimes reflect the change in the contribution of $X_D^{-}$, which undergoes thermal activation as a result of $XX^-$ decomposition in agreement with our simulations. We propose that dark trions act as reservoirs for both complexes, which is a key point in our self-consistent model describing the PL kinetics in the WS$_2$ monolayer.
   
Our study links the phenomenological description of exciton luminescence spectra with a quantitative kinetics model applicable to the ''darkish'' 2D materials based on transition metal dichalcogenides. Our results evidence that 2D darkish exhibiting a set of different excitonic states present a rich platform for studying many-body physics, being also important for designing photonic devices.

\section*{Author contributions}
Optical measurements: A.I.G., A.I.V. Model: A.I.V., M.M.G. Data simulation: A.I.V. Research initiation and supervision: T.V.S.

\section*{Acknowledgments}
The authors thank I. A. Eliseyev for the help in the sample preparation.

\bibliography{rsc}

\appendix

\section{Sample preparation}
Atomically thin WS$_2$ flakes were exfoliated from the naturally negatively doped bulk crystal (production of HQ Graphene) and transferred by a conventional method onto a 285-nm-thick SiO$_2$ thin film on a degenerately doped Si substrate. Monolayer-thickness regions were determined by comparing PL and Raman spectra measured at room temperature. The monolayer region is characterized by the absence of the PL line associated with the indirect exciton transition and the absence of the low-frequency layer-breathing mode ($\sim$40 cm$^{-1}$), which requires the presence of a second layer.

We did not use hBN encapsulation, which can cause sample deformation, potentially altering the exciton spectrum \cite{Eliseyev2021,XuHan2019}. Instead, we immediately transferred the sample after preparation to a pumped cryostat for measurements. A comparison of the measured exciton spectra with published data on the electrically controlled electron density \cite{Chernikov2015} shows that the electron density in the studied monolayer is  several $10^{12}$ cm$^{-2}$ that is markedly less than a value in the middle of $10^{13}$ cm$^{-2}$, at which complete ionization of excitons can occur.

\section{Optical measurements}
Optical measurements of WS$_2$ monolayers were performed in the temperature range from 8 to 300~K. The sample was mounted in a ST-500-Attocube cryostat (Janis) equipped with a three-axis piezoelectric positioning stage located in the cold zone of the cryostat, providing a positioning accuracy of approximately 20~nm. Excitation was provided by a femtosecond pulsed parametric amplifier operating at a wavelength of 525~nm with a repetition rate of 78.3~MHz (Avesta TOPOL). The excitation beam was focused onto the sample surface using a 100x apochromatic objective lens (Mitutoyo Plan Apo, NA~=~0.7), resulting in a laser spot diameter of approximately 1~$\mu$m. Under these conditions, the excitation power density on the sample surface was estimated to be $\sim$300~W/cm$^2$. The PL signal was collected by the same objective in a confocal backscattering geometry and directed to the entrance slit of an SP-2500 spectrometer equipped with a cooled PyLoN CCD detector (Princeton Instruments). Scattered laser radiation was suppressed using a band-pass interference filter. TRPL measurements were performed using superconducting single-photon detectors (Scontel) with a temporal resolution of approximately 30~ps.

\section{Origin of the $L_R$ peak}
\begin{figure*}[t]
	\centering
	\includegraphics[width=\linewidth]{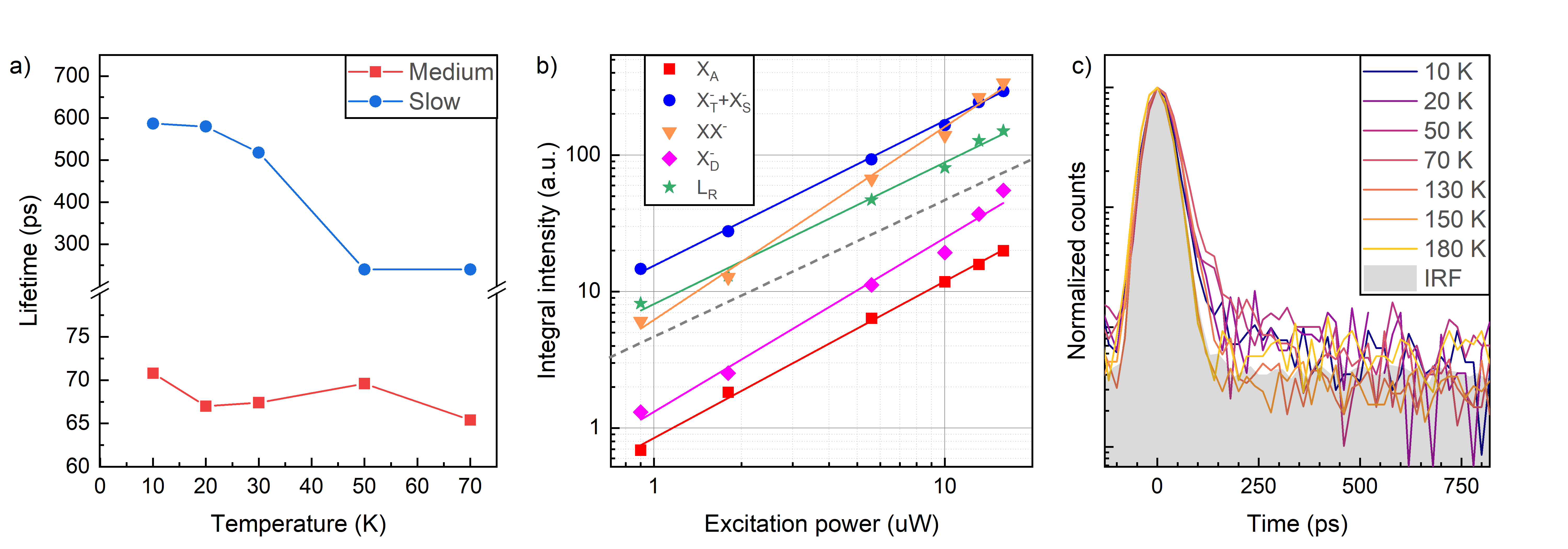}
	\caption{a) TRPL data for the $L_R$ peak: blue (red) symbols correspond to the slow (middle) lifetime component. b) Laser power dependence of integrated PL intensity of the excitonic complexes under study. Symbols represent the experimental data, while lines represent a power law fit. Grey dashed line represents strictly linear law ($\alpha =1$). c) TRPL curves for the $X_A$ line emission at different temperatures. Shaded area represents IRF.}
	\label{SM_fig}
\end{figure*}

PL line, noted as $L_R$ in Figure \ref{fig1}, exists at all used excitation intensities and can be accurately spectrally separated at temperatures below 90 K. TRPL data for the $L_R$ line are presented in Figure \ref{SM_fig}a. Radiative lifetime for the fast component (not shown in Figure \ref{SM_fig}a) is fixed at 10 ps, and integral impact of the fast component is higher than 50\% at all temperatures. Taking into account near-linear power dependence of $L_R$ integral PL intensity (see section S2), only slow component with a relative impact around 10\% can be related to defect-assisted recombination. The main PL signal of the $L_R$ peak can be attributed to phonon replicas of $X_D^-$ dark trion. Energy separation between $L_R$ and $X_D^-$ peaks is around 20 meV, which nearly coincides with energy of $K_3$ and $\Gamma_5$ valley phonons \cite{He2020}.

\section{Fitting parameters for temperature dependencies of peak energies}
To fit the temperature dependence of PL peaks energies, we use standard cotangent relation from the Ref.~\cite{PhononCotangent} in the next form:
\[E(T)=E_0-S\Delta\coth\left(\frac{\Delta}{2kT}\right),\]
where $E_0$ is the energy position at $T=0$, $\Delta$ is the is an average phonon energy, $S$ is a parameter governing exciton-phonon coupling, and $k$ is the Boltzmann constant. Fitting parameters, used to reconstruct experimental data in Fig.~\ref{fig2}, are collected in Tab.~\ref{cot_pars}.

\begin{table}[h]
    \centering
    \begin{tabular}{|c|c|c|c|}
    \hline
              & $E_0$, eV & $\Delta$, meV & $S$ \\
    \hline
        $X_A$ & 2.0991 & 17.7 & 2.3\\
    \hline
        $X^-$ & 2.0573 & 15.7 & 2.5\\
    \hline
        $XX^-$ & 2.0363 & 29.7 & 4.3 \\
    \hline
        $X_D^-$ & 2.0288 & 15.6 & 3.3 \\
    \hline
    \end{tabular}
    \caption{Fitting parameters for temperature dependence of peak PL energy of excitons under investigation.}
    \label{cot_pars}
\end{table}

\section{Power dependence of integrated PL intensity}
Dependence of integrated PL intensity on excitation power for all studied excitonic complexes is presented with markers in Figure \ref{SM_fig}b in double logarithmic scale. Solid lines show linear fit, corresponding to power law fit of the form  $AP^\alpha$, where $A$ is a scaling parameter, $P$ is excitation power, $\alpha$ is exponential coefficient. From the fit we obtained the following results for $\alpha$ coefficients:  1.15$\pm$0.03 ($X_A$), 1.05$\pm$0.02 ($X^-$), 1.41$\pm$0.05 ($XX^-$), 1.15$\pm$0.05 ($X_D^-$), 1.04$\pm$0.04 ($L_R$). As can be seen, only the $XX^-$ feature shows distinctive superlinear dependence on excitation power.

\section{TRPL measurements and fitting procedure}
To minimize the impact of peaks overlapping, we used band-pass filters to select the narrow ($\sim$5 meV) spectral window in the center of each line during data acquisition. 

To obtain a lifetime fit curve based on TRPL experimental data, we convolve exponential decay functions with impulse response functions (IRFs) of single photon detectors.
Exponential decay functions $f_i(t)$ with lifetimes $\tau_i$ were taken as \(f_i(t)=A_i\exp(-t/\tau_i)\cdot\Theta(t)\), where $A_i$ is a magnitude of individual component, $\Theta(t)$ is the Heaviside step function.
For Gaussian-like IRF with time resolution $\sigma$ we obtained analytical fitting function:
\[I_{\text{fit}}(t)=I_0+\sum_i \frac{A_i}{2}\exp\left(\frac{\sigma^2-2\tau_it}{2\tau_i^2}\right)\text{erfc}\left(\frac{\sigma^2-\tau_it}{\sqrt{2}\tau_i\sigma}\right),\]
where $I_0$ (dark counts of detectors) determined by acquisition time, while time resolution $\sigma$ determined from fitting of detectors response on reflected femtosecond laser pulses, erfc($x$) - complementary error function. Index $i$ in sum runs through lifetime components designations (fast, slow or middle), described in the main text. For accurate determination of fast component with lifetimes much shorter than IRF, during fitting we fixed fast component lifetime at 10 ps.

\section{TRPL data for $X_A$ line}
Figure \ref{SM_fig}c presents decay curves for neutral bright exciton $X_A$. For every studied temperature radiative lifetime of the $X_A$ is shorter than the time resolution of detectors, and curves become almost indistinguishable from IRF.

\section{Final form of rate equations}

The system of rate equations in explicit form is presented in Eq.(\ref{system}). Here, $\tilde\gamma(\tau)$ is the effective radiative lifetime accounting states inside the radiative light cone: \(\tilde\gamma(\tau_i)=\tau_i^{-1}2E_0/(3kT)\), where $E_0$ and $\tau_i$ are microscopic parameters determined from first-principles calculations \cite{Palummo2015,Robert2016}. The factor \(n(\Delta_i) = (\exp(\Delta_i/kT)-1)^{-1}\) is the Bose-Einstein number of phonons with energy $\Delta_i$ and stands for dark-to-bright transitions\cite{Labeau2003} and for exciton-trion conversion\cite{Ayari2020}. Factor $\exp(-E_b/kT)$ controls temperature dependent dissociation of the charged biexciton $XX^-$. 

For quantitative comparison we introduce the following microscopic parameters: $\gamma_{XT}$ determines \(X_A\leftrightarrow X^-\) and \(X_D\leftrightarrow X_D^-\) conversion processes rates, $\gamma_{AD}$ controls dark-bright interplay in \(X_A\leftrightarrow X_D\) and \(X^-\leftrightarrow X_D^-\) pairs and $\gamma_{AXT}$ describes formation and dissociation of the $XX^-$ complex. Parameters $\Delta_i$ designed for the energy separations between $X_A$ and $X^-$ ($\Delta_{AT}$), $X_A$ and $X_D$ ($\Delta_{AD}$), $X^-$ and $X_D^-$ ($\Delta_{TD}$), $X_D$ and $X_D^-$ ($\Delta_{DAT}$). All parameters with used values presented in the Table \ref{tabS5} below. Energy separation between $X_D$ and $X_D^-$ lines derived from other values as \(\Delta_{DAT} = \Delta_{AT} + \Delta_{DT} - \Delta_{AD}\). Boundary conditions for this system takes into account zero initial population for all complexes except $X_A$, whose initial population is equal to the surface density of excited electron-hole pairs. For actual experimental conditions (laser spot size 3$\mu$m, average excitation power 10$\mu$W, excitation wavelength 525 nm and delay between laser pulses 12.77 ns, relative absorbance of WS$_2$ monolayer 0.02 at 525 nm) we evaluated the number of excited electron-hole pairs around 9.6$\cdot$10$^{10}$ cm$^{-2}$. Doping level $n_e$ in WS$_2$ monolayer (n-type) was estimated as 10$^{12}$ cm$^{-2}$. Final (for $t\rightarrow\infty$) population for all complexes is set to zero. 

\begin{figure*}[t]
\caption{System of numerically solved rate equations in explicit form.}
\begin{equation}\label{system}\left\{
\begin{split}
    \frac{d}{dt}N_A(t)=&-\gamma_{XT}(n(\Delta_{AT})N_A(t)-(n(\Delta_{AT})-1)N_T(t))+\gamma_{AD}((n(\Delta_{AD})-1)N_D(t)-n(\Delta_{AD})N_A(t))-\\
    &-\gamma_{AXT}N_A(t)N_{DT}(t)+\gamma_{AXT}\exp(-E_b/kT)N_{XX}(t)-\tilde\gamma(\tau_A)N_A(t)\\
    \frac{d}{dt}N_T(t)= &\ \ \gamma_{XT}(n(\Delta_{AT})N_A(t)-(n(\Delta_{AT})-1)N_T(t))+\gamma_{AD}((n(\Delta_{TD})-1)N_{TD}(t)-n(\Delta_{TD})N_T(t))-\\
    &-\gamma_{AXT}N_T(t)N_D(t)+\gamma_{AXT}\exp(-E_b/kT)N_{XX}(t)-\tilde\gamma(\tau_T)N_T(t)\\
    \frac{d}{dt}N_D(t)=&-\gamma_{XT}(n(\Delta_{DAT})N_D(t)-(n(\Delta_{DAT})-1)N_{DT}(t))-\gamma_{AD}((n(\Delta_{AD})-1)N_D(t)-n(\Delta_{AD})N_A(t))\\
    &-\gamma_{AXT}N_T(t)N_D(t)+\gamma_{AXT}\exp(-E_b/kT)N_{XX}(t)-\frac{1}{\tau_D}N_D(t)\\
    \frac{d}{dt}N_{DT}(t)=&\ \ \gamma_{XT}(n(\Delta_{DAT})N_D(t)-(n(\Delta_{DAT})-1)N_{DT}(t))-\gamma_{AD}((n(\Delta_{TD})-1)N_{DT}(t)-n(\Delta_{TD})N_T(t))+\\
    &+(\gamma_{AXT}\exp(-E_b/kT)+\tilde\gamma(\tau_A))N_{XX}(t)-\gamma_{AXT}N_A(t)N_{DT}(t)-\frac{1}{\tau_{DT}}N_{DT}(t).\\
    \frac{d}{dt}N_{XX}(t)=&\ -\left(4\gamma_{AXT}\exp(-E_b/kT)+\tilde\gamma(\tau_A)\right)N_{XX}(t)+2\gamma_{AXT}(N_A(t)N_{DT}(t)+N_T(t)N_D(t))\\
\end{split}\right.
\end{equation}

\end{figure*}

\begin{table*}
    \centering
    \caption{Microscopic parameters, used in simulations. Values marked with * were taken from experimental PL spectra, cited values were taken from literature.}
    \begin{tabular}{|c|c|c|c|c|c|c|c|c|c|c|c|}
    \hline
         $\tau_A$, ps & $\tau_T$, ps &  $\tau_D$, ps &  $\tau_{DT}$, ps &  $\gamma_{AD}$, ps$^{-1}$ &  $\gamma_{AT}$, ps$^{-1}$ &  $\gamma_{AXT}\cdot n_e$, ps$^{-1}$ & $\Delta_{AT}$, meV & $\Delta_{AD}$, meV & $\Delta_{TD}$, meV & $E_b$, meV\\
         \hline
         0.19 \cite{Palummo2015}& 0.5\cite{Robert2016} & 700\cite{Lagarde2024} & 285 & 0.1\cite{Lagarde2024} & 2\cite{Ayari2020} & 0.6 & 39* & 45 & 33* & 23*\\
         \hline
    \end{tabular}
    
    \label{tabS5}
\end{table*}

\nocite{*}

\end{document}